\documentclass[a4paper,fleqn,usenatbib]{mnras}

\usepackage{mathptmx}

\usepackage[T1]{fontenc}
\usepackage{ae,aecompl}

\usepackage{graphicx}   
\usepackage{amsmath}    
\usepackage{amssymb}    

\usepackage{comment}
\includecomment{versiona}

\usepackage[usenames]{color}

\usepackage[normalem]{ulem}

\newcommand{\be}{\begin{equation}}
\newcommand{\ee}{\end{equation}}
\newcommand{\bi}{\begin{itemize}}
\newcommand{\ei}{\end{itemize}}

\title[Modified gravity and globular clusters]{Testing modified gravity with globular clusters:  the case of NGC 2419}

\author[Llinares, C.]{
Claudio Llinares,$^{1,2}$\thanks{E-mail: claudio.llinares@durham.ac.uk}
\\
$^{1}$Institute for Computational Cosmology, Department of Physics, Durham University, Durham DH1 3LE, U.K.\\
$^{2}$Institute of Theoretical Astrophysics, University of Oslo, 0315 Oslo, Norway
}

\date{Accepted XXX. Received YYY; in original form ZZZ}

\pubyear{2017}

\begin{document}
\label{firstpage}
\pagerange{\pageref{firstpage}--\pageref{lastpage}}
\maketitle

\begin{abstract}
The dynamics of globular clusters has been studied in great detail in the context of general relativity as well as with modifications of gravity that strongly depart from the standard paradigm such as MOND.  However, at present there are no studies that aim to test the impact that less extreme modifications of gravity (e.g. models constructed as alternatives to dark energy) have on the behaviour of globular clusters.  This Letter presents fits to the velocity dispersion profile of the cluster NGC 2419 under the symmetron modified gravity model.  The data shows an increase in the velocity dispersion towards the centre of the cluster which could be difficult to explain within general relativity.  By finding the best fitting solution associated with the symmetron model, we show that this tension does not exist in modified gravity.  However, the best fitting parameters give a model that is inconsistent with the dynamics of the Solar System.  Exploration of different screening mechanisms should give us the chance to understand if it is possible to maintain the appealing properties of the symmetron model when it comes to globular clusters and at the same time recover the Solar System dynamics properly.
\end{abstract}




\section{Introduction}

With few exceptions, the majority of the current data shows that globular cluster are not immersed in dark matter halos as usually considered to be the case with galaxies.  This unique characteristic reduces the number of parameters required to explain their dynamics and thus, makes them ideal laboratories to test gravity.  In the context of MOND, several authors have shown the power of these objects to give constraints on standard and modified gravity (MG) theories \citep{2011ApJ...738..186I, 2012MNRAS.419L...6S, 2012MNRAS.422L..21S, 2011ApJ...743...43I, 2014ApJ...783...48D, 2009AJ....137.4586J, 2010A&A...509A..97G}.  However, this power has not been exploited yet in the context of modifications of gravity that are intended to explain the accelerated expansion of the Universe.  The aim of this Letter is to fill this gap by studying the impact of a particular gravitational model on the velocity dispersion profile of a specific globular cluster.  Two complementary aspects of the problem will be studied.  Firstly, we will investigate if MG can improve the fits of velocity dispersion profiles of clusters in cases where standard gravity is in tension with the observations.  Secondly, we want to understand if globular clusters can be useful in constraining the parameter space of modified gravity theories.

This Letter will deal with a particular gravitational model: the symmetron model \citep[][]{2005PhRvD..72d3535P, 2008PhRvD..77d3524O, 2010PhRvL.104w1301H}.  Nevertheless, the results can be easily extrapolated to any model in which a fifth force is present.  The model includes a screening mechanism, which means that it has a general relativity (GR) limit in high density regions and thus, it is able to pass Solar System tests.  At the same time the model includes a fifth force in the dynamics of the systems on galactic and cosmological scales.  As mentioned previously the model is not intended to give an explanation for the dark matter content of the universe, but belongs to a family of theories that give a slight modification of gravity on the scales of galaxies and beyond and is intended to provide an explanation for the accelerated expansion of the Universe.

The target that will be used to carry our analysis is the Milky Way globular cluster NGC 2419, which is located about 80 kpc from the galactic center.  From the modified gravity point of view, this large distance is an important feature because, otherwise, the dynamics of the system would be dominated by external field effects produced by the galaxy itself.  In other words, the scalar field will be screened by our own galaxy for small galacto-centric distances and thus, MG effects would not be measurable.  The second reason to study this cluster in particular is that there are good measurements of its luminosity and velocity dispersion profiles \citep{2011ApJ...738..186I}, which will be necessary for our analysis.

Fits to these observations were provided by \cite{2011ApJ...738..186I} and \cite{2013MNRAS.428.3648I} in a GR context.  There it was shown that GR underestimates the central velocity dispersion when the velocities are assumed to be isotropic.  The tension can be reduced by including anisotropic velocity dispersions or a dark matter component \citep{2011ApJ...738..186I, 2013MNRAS.428.3648I}.  In this Letter, we explore the possibility that MG can help to improve the fit without invoking extra astrophysical effects.  As we do not know what the impact of modified gravity will be in the predictions, we will assume the simplest model for the cluster and thus keep the isotropy assumption.  A more detailed analysis including anisotropic velocities is left for future work.

The Letter is organized as follows:  Section \ref{section:symmetron} summarizes a few aspects of the MG model that we consider.  Section \ref{section:analysis} describes the methodology used to obtain the best fit models.  The results and conclusions are presented in Sections \ref{section:results} and \ref{section:conclusions}.

\section{The symmetron model}
\label{section:symmetron}

The symmetron model belongs to the family of scalar-tensor theories and is defined by the following action:
\be
S = \int \sqrt{-g} \left[ \frac{M_p^2}{2} R - \frac{1}{2}\nabla^a\phi \nabla_a \phi - V(\phi)\right] d^4x + S_M(\tilde{g}_{ab}, \psi) \ ,
\label{symm-action}
\ee
where the scalar field $\phi$ is an extra degree of freedom which will be responsible for the fifth force introduced by the model.  The Einstein and the Jordan frame metrics ($g_{ab}$ and $\tilde{g}_{ab}$) are related according to
\be
\tilde{g}_{ab} = A^2(\phi) g_{ab},
\ee
and the free potential and the conformal factor have the following forms:
\begin{align}
\label{potential}
V(\phi) &= -\frac{1}{2}\mu^2\phi^2 + \frac{1}{4}\lambda\phi^4 + V_0 \\
\label{conformal_factor}
A(\phi) &= 1 + \frac{1}{2}\left(\frac{\phi}{M}\right)^2 \ ,
\end{align}
where $\mu$ and $M$ are mass scales, $\lambda$ is a dimensionless constant and $V_0$ is tuned to match the observed cosmological constant.  The analysis is made assuming a Minkowski background.

For numerical reasons, it is convenient to work with a dimensionless scalar field $\chi \equiv \phi/\phi_0$, where $\phi_0$ is the expectation value that corresponds to $\rho=0$:
\be
\phi_0 = \frac{\mu}{\sqrt{\lambda}}.
\ee
It also helps to relate the three free parameters of the model, $(\mu, \lambda, M)$, to the Compton wavelength of the scalar field at $\rho=0$,
\be
\lambda_0 = \frac{1}{\sqrt{2}\mu} \ , 
\label{def_lambda0}
\ee
a dimensionless coupling constant, 
\be
\beta = \frac{\phi_0 M_{\mathrm{pl}}}{M^2} \ ,
\label{def_beta}
\ee
and the expansion factor at the time of the symmetry breaking,
\be
a_{\mathrm{SSB}}^3 = \frac{\rho_0}{\rho_{\mathrm{SSB}}} = \frac{\rho_0}{\mu^2 M^2} \ , 
\label{def_assb}
\ee
where $\rho_0$ is the background density of the Universe at redshift $z=0$.  This last parameter fixes the moment in the history of the Universe in which the MG effects are activated in the background.  For expansion factors smaller than $a_{\mathrm{SSB}}$ (or redshifts larger than its corresponding $z_{\mathrm{SSB}}$ value), the scalar field is completely screened and thus, there is no MG effect.  The opposite happens when the expansion factor becomes larger that $a_{\mathrm{SSB}}$.  The field equation for the dimensionless scalar field $\chi$ is then
\be
\nabla^2\chi = \frac{1}{2\lambda_0^2}\left[\left(\frac{\rho}{\rho_{\mathrm{SSB}}} - 1\right)\chi + \chi^3 \right] = \frac{d}{d\chi}V_{\mathrm{eff}}(\chi), 
\label{eq_motion_chi}
\ee
where $\rho$ is the Jordan frame matter density.  From the form of $V_{\mathrm{eff}}(\phi)$, one can see that the expectation value of the scalar field vanishes at high matter densities, setting the conformal factor $A(\phi)$ to unity, decoupling the scalar from the matter and thus screening its effects.  For low densities, the potential develops two minima away from zero and thus, the scalar field can adopt values different to zero and therefore, change the dynamics of the systems.

The Solar System constraints on the model parameters are the following:
\begin{align}
\label{allowed_models}
\frac{\lambda_0}{a_{\mathrm{SSB}}^{3/2}} & \lesssim \frac{10^{-3}c}{\sqrt{6\Omega_{\mathrm{m}}}{H_0}}\\
\label{constraint_2}
z_{\mathrm{SSB}} & \lesssim 150, 
\end{align}
where $c$ is the speed of light, $\Omega_{\mathrm{m}}$ is the background matter density at redshift $z=0$ and $H_0$ is the Hubble constant.  The first constraint \citep{2010PhRvL.104w1301H, 2014PhRvD..90l4041L} is based on the post-Newtonian parameter $\gamma$ that was measured by the Cassini experiment \citep{Cassini}.  The second constraint is a consequence of the fact that the derivation of Eq. \ref{allowed_models} assumes that the Milky Way screens the scalar field at the position of the Sun.  A proxy for this condition can be determined by comparing the density of symmetry breaking $\rho_{\mathrm{SSB}}$ with the density of the galaxy at the position of the Solar System, which has been estimated, for instance, by \cite{2013A&A...549A.137I}.  Model parameters for which $\rho_{\mathrm{SSB}}$ is larger than the local density will not screen the Solar System and thus will be excluded.  This condition is summarized by Eq. \ref{constraint_2}.

\section{Analysis}
\label{section:analysis}

The analysis consists in finding the best fit model to the observed velocity dispersion profile.  While in this Letter we are only interested in the velocity distribution, we will also make use of luminosity information.  The data used for the fitting was presented in \cite{2011ApJ...738..186I}.  For the velocity dispersion profile we used the sample A+B presented in that work.  In order to construct the $\chi^2$ quantities that will be used for the fitting, we need a prediction for luminosity and velocity profiles for both standard and modified gravity.  In the Newtonian case, we assumed a King spherical isotropic model \citep{1966AJ.....71...64K}.  The density and velocity dispersion profiles were calculated as integrals of the King phase space distribution function.  The normalisation of the fields as well as details of the calculation follow closely the approach described by \cite{1966AJ.....71...64K}.

For the symmetron case we need to construct a self-consistent model given by density and velocity distributions that are compatible with this particular gravitational model.  For the density distribution, we choose the best fit King model that was obtained with Newtonian gravity.  In this case, the King model does describe the dynamics of the system, but is only a way of parametrising the density distribution.  The model is then made self-consistent under modified gravity by solving the Jeans equation (see for instance \citet{1987gady.book.....B}) including the fifth force associated to the symmetron model:
\be
\frac{d\sigma^2}{dR} + \frac{1}{\rho} \frac{d\rho}{dR} \sigma^2 + \frac{dW}{dR} = 0, 
\ee
where $dW/dR$ is the total force:
\be
\frac{dW}{dR} = \frac{d\Phi}{dR} + 6 H_0^2 \Omega_0 \frac{\lambda_0^2\beta^2}{a_{\mathrm{SSB}}^3} \chi \frac{d\chi}{dR}, 
\ee
$\sigma$ is the velocity dispersion profile and $\Phi$ is the gravitational potential, which is solution of the standard Poisson's equation.  The required solution for $\chi(R)$ was obtained by solving Eq. \ref{eq_motion_chi} numerically.  The variables used during the integration were again closely related to those described by \cite{1966AJ.....71...64K}.

The Newtonian model contains five free parameters:  the central value of the potential $w_0$, which is normalised according to \citet{1966AJ.....71...64K}, the mass of the cluster $M$, the core radius $r_{\mathrm{c}}$, the distance to the cluster $d$ and the mass-to-light ratio $\Upsilon$.  In order to simplify the fitting procedure, we fixed the distance of the cluster to the value that was measured with RR Lyrae variables by \citet{2007ApJ...667L..61R}:  $d=83.2$ kpc.  The mass to light ratio was fixed by calculating the total luminosity from the observed luminosity profile and taking into account the mass of the cluster.  After making these simplifications, we end up with only three free parameters: $w_0$, $r_{\mathrm{c}}$ and $M$.  The fit was made using a grid in the parameter space to find the smallest $\chi^2$ value.  

In the MG case, we need to fit also the three free parameters of the gravitational model $(\lambda_0, a_{\mathrm{SSB}}, \beta)$.  As previously stated, this fit was made assuming the luminosity profile obtained from the best Newtonian model.  This fixes the parameters $w_0$ and $r_{\mathrm{c}}$.  The luminosity is independent of the mass $M$ (changes in $M$ can be compensated by opposite changes in $\Upsilon$), so we keep this parameter free for the MG analysis.  This is important since the increase in the velocity dispersion produced by the modifications to gravity can be compensated in part by reducing the mass of the cluster.  We found that this can help to find a much better overall fit to the velocity profile.  To summarise, the MG fit to the velocity profile was made assuming four free parameters: $(M, \lambda_0, a_{\mathrm{SSB}}, \beta)$.  In order to take into account the Solar System constraints, we substitute the parameter $a_{\mathrm{SSB}}$ by a new parameter $\xi$.  The relationship between these parameters was obtained from Eq. \ref{allowed_models} and is given by:
\be
\label{transformation}
a_{\mathrm{SSB}} = \frac{1-C\lambda_0^{2/3}}{1-C\lambda_{0,\textrm{min}}^{2/3}}\xi + \frac{C(\lambda_0^{2/3}- \lambda_{0,\textrm{min}}^{2/3})}{1-C\lambda_{0,\textrm{min}}^{2/3}}, 
\ee
where we assumed that the maximum value that can be reached by $a_{\mathrm{SSB}}$ is 1 (i.e. today) and $C$ is defined by $C=[(H_0\sqrt{6\Omega_{\mathrm{m}}})/(10^{-3}c)]^{2/3}$.  The second constraint on the model parameters given by Eq. \ref{constraint_2} will be taken into account \textit{a posteriori}. The transformation given by Eq. \ref{transformation} maps the allowed region of the parameter space (according to Eq. \ref{allowed_models}) to an area that lies inside a rectangle.  The lowest value of $\xi$ used during the fitting procedure corresponds to the minimum value of $a_{\mathrm{SSB}}$ allowed by the solar system given the minimum value that was chosen for $\lambda_0$.  The upper boundary of the domain is given by $\xi_{\mathrm{max}}=a_{\mathrm{SSB,max}}=1$.

There is one final constraint that was included in the parameter space.  \citet{2012MNRAS.423..844B} provided limits on the mass to light ratio which are based on stellar evolution models: $1.2<\Upsilon<1.7$.  Assuming the observed total luminosity and a fixed value for the distance, we can translate these constraints on $\Upsilon$ to constraints on the total mass of the cluster: $4.9 \times 10^5 < M/M_{\odot} < 6.9 \times 10^5$.  These constraints were taken into account when minimising $\chi^2$.  However, we found that under these conditions, Newtonian gravity gives a very poor fit to the data.  For this reason, we increased the limits on the mass only for this case (see \citet[][]{2012MNRAS.423..844B} for a discussion on this tension).  Thus, we will compare the most favourable situation for Newtonian gravity with the most restrictive for MG.  Under these conditions, we will find that MG can still give a better fit to the velocity data.

\begin{figure}
  \begin{center}
    \includegraphics[width=0.5\textwidth]{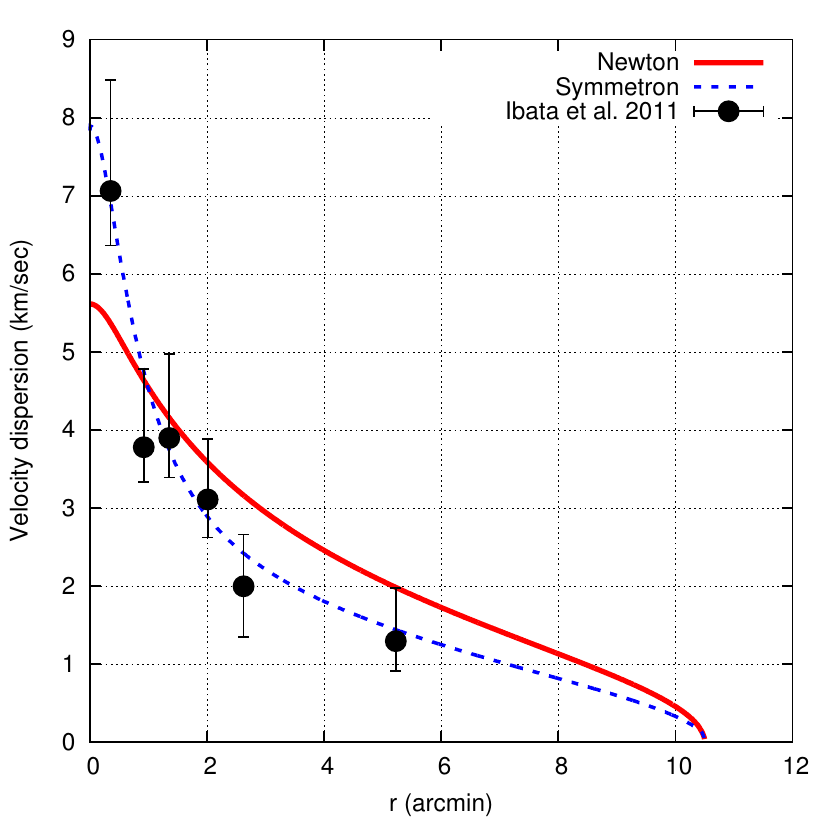}
    \caption{Observed velocity dispersion profile compared to the best fit models for Newtonian gravity and the symmetron model.  The observed data is from \citet{2011ApJ...738..186I}.} 
    \label{fig:fits}
  \end{center}
\end{figure}

\begin{figure*}
  \begin{center}
    \includegraphics[width=0.95\textwidth]{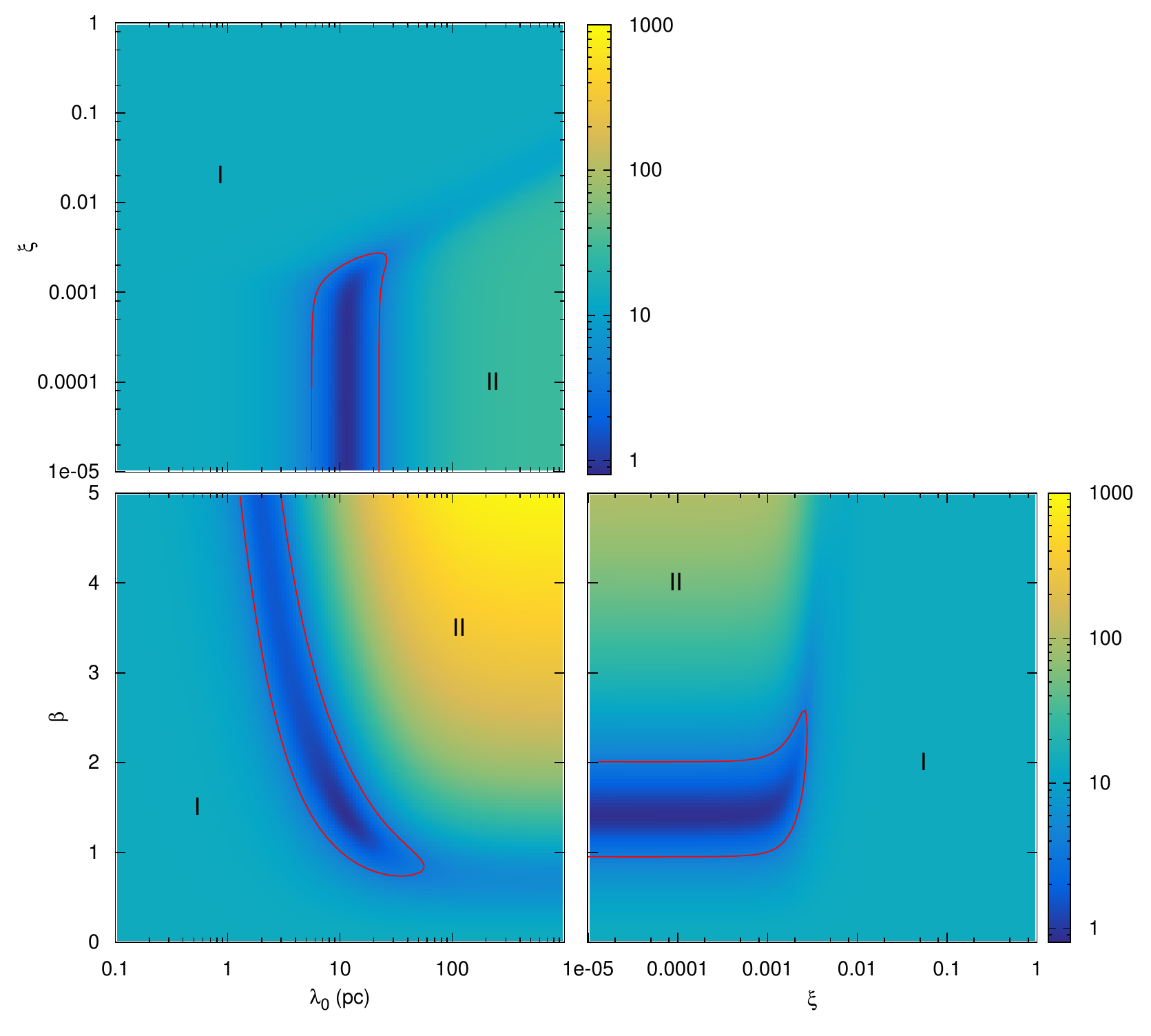}
    \caption{Cuts in the parameter space of the symmetron model with planes that pass through the minimum $\chi^2$ value.  The continuous line corresponds to the value of reduced $\chi^2$ that was obtained with Newtonian gravity.  Models lying in the region inside this curve give an improved fit with respect to Newtonian gravity.  See section \ref{section:results} for further explanation.}
    \label{fig:maps_chi}
  \end{center}
\end{figure*}

\section{Results}
\label{section:results}

Fig. \ref{fig:fits} shows the velocity dispersion data points and the best fit solutions obtained from the Newtonian and MG models.  The Newtonian model fails to explain the large increase in the velocities towards the centre.  On the other hand, the fifth force can boost the velocities in the centre of the cluster helping to explain the innermost point. A renormalisation of the mass of the cluster can shift the whole profile to lower values, providing also a better fit to the outermost point.  The best Newtonian and MG model parameters are $(w_0, M, r_{\mathrm{c}})=(7.25, 1.027\times 10^6 ~ \mathrm{M}_{\odot}, 6.33 ~\mathrm{pc})$ and $(M, \lambda_0, a_{\mathrm{SSB}}, \beta) = (5.21 \times 10^5 \mathrm{M}_{\odot}, 11.5 ~\mathrm{pc}, 0.000236, 1.41)$ respectively.

The minimum values of the reduced $\chi^2$ associated with the velocity dispersion profile are 3.72 and 0.87 for Newtonian and MG models respectively.  Here we take into account the 6 data points and the 3 and 4 parameters used for the fit in each case.  While in the MG model case we used one more free parameter, the $\chi^2$ values confirm what can be seen by eye in Fig. \ref{fig:fits}: the MG model provides a better fit to the velocity profile.  In addition to this, the tension that exists in the Newtonian gravity model related to the mass of the cluster disappeared.

The distribution of $\chi^2$ values in the parameter space of the symmetron model presents a complex structure, which is shown in Fig. \ref{fig:maps_chi}.  The three panels correspond to three cuts in the parameter space with planes that pass through the absolute minimum of $\chi^2$.  The colours represent the reduced $\chi^2$ associated with the symmetron fit and the red contour corresponds to the value of reduced $\chi^2$ that was obtained with Newtonian gravity.  All the model parameters that lie inside this contour give an improved fit with respect to Newtonian gravity.  The rest of the parameter space appears divided into two regions.  The regions marked with ``I'' correspond to the Newtonian limit of the model, where the Newtonian limit must be understood as defined by the present data.  All of the model parameters lying in this region give a negligible fifth force and thus, provide a velocity dispersion profile which is equal to the one predicted by Newtonian gravity.  Note that the $\chi^2$ value in this region is higher than the $\chi^2$ value for the Newtonian model one only because the MG model has one more free parameter.  The region ``II'' corresponds to models for which the fifth force becomes too large and ruins the fit at all radii, giving much larger $\chi^2$ values. 

The observations can not give an upper bound to the coupling constant $\beta$.  The reason for this is that the data does not provide an upper bound for the central velocity dispersion.  The models with large $\beta$ correspond to models that provide a good fit to all of the observed points and at the same time have a very large velocity dispersion in the region that lies between the centre of the cluster and the first data point.  Observations of the turnover in the central velocity dispersion profile will be necessary to determine an upper bound to the coupling constant $\beta$.

While the fits presented here show that MG can give a much better explanation of the inner velocity dispersion profile, there is the problem that $\xi$ has to be below $0.005$ for this to happen.  This corresponds to a redshift of symmetry breaking above the limit given by Eq. \ref{constraint_2}, and thus it is ruled out by Solar System observations.  So the cluster does not rule out MG, but the regions of the parameter space that provide improved fits to its internal dynamics are ruled out by Solar System observations.

\section{Conclusions}
\label{section:conclusions}

We have studied, for the first time, the impact of alternatives to dark energy on the velocity dispersion profile of the globular cluster NGC 2419.  Our aim was to understand both how these particular gravity models affect the velocity dispersion profile and what these observations can say about the parameter space of MG models.

The globular cluster has a large galacto-centric distance and thus, it is expected that its internal dynamics will not be affected by the presence of the Milky Way.  Furthermore, the cluster is expected not to have a dark matter component.  This makes this object an ideal laboratory to study non-linear theories for gravity which include screening mechanisms.  Observations of the velocity dispersion profile \citep{2011ApJ...738..186I} show an increase of the velocity towards the centre of the globular cluster which may be difficult to explain with standard gravity when the velocities are assumed to be isotropic.  Here we present fits to these observations within the context of the symmetron MG model.  The model is a scalar tensor theory that includes a screening mechanism which allows it to pass Solar System tests; at the same time it gives rise to a fifth force that could affect the dynamics of globular clusters.  We found that the presence of the scalar field can boost the velocities in the centre of the globular cluster and thus provides a much better fit than with Newtonian gravity.  Furthermore, we found that the mass of the globular cluster that corresponds to the best MG model is consistent with the observed mass-to-light ratio, which is not the case for Newtonian gravity.

The typical model parameters used in cosmological studies of the symmetron model \citep{2012ApJ...748...61D, 2014A&A...562A..78L, 2014PhRvD..89h4023L, Hagala1} correspond to a scalar field that is switched-on at late times ($z_{\mathrm{SSB}}\sim 1$) and a vacuum Compton wavelength $\lambda_0$ of the order of 1 Mpc/$h$.  In contrast, we found here a scalar field that is activated at early times ($z_{\mathrm{SSB}} \gg 1$) and has a much smaller vacuum range $\lambda_0$.  Note that this new regime has been studied by \cite{2014PhRvD..90l4041L}, where it was shown that a symmetron field with these values can give rise to high energy cosmological domain walls which should be observable in the cosmological data.  While the phenomenology associated with the symmetron field is very rich in this regime, the required model parameters do not allow the screening mechanism to suppress the scalar field in the inner part of the galaxy and thus, Solar System tests can not be passed.  Different screening mechanisms (e.g. chameleon, Vanshtein, etc.) should be studied before accepting that this result can be generalized to all MG theories.

The effects of MG are maximal at the centre of the globular cluster.  Increasing the number of data points in this region will be crucial to give an upper bound to the coupling constant of the theory.  Future advances in astronometry such as the incoming Micado instrument in ELT will make possible to obtain accurate measurements of the orbital structure of globular clusters.  This will in turn make possible to develop novel tests for the symmetron and similar models.

\section*{Acknowledgements}

I am very grateful to Yashar Akrami, Miguel Quartin, Hans Kristian Heriksen, H{\aa}kon Dahle, David Mota, Carlton Baugh, Baojiu Li, Shaun Cole and David Alexander for helpful discussions and carefully reading the manuscript.  I acknowledge support from the Research Council of Norway through grant 216756 and STFC consolidated grant ST/L00075X/1 \& ST/P000541/1.

\bibliographystyle{mnras}
\bibliography{references}


\bsp    
\label{lastpage}

\end{document}